\documentclass[]{aero}

\usepackage{xcolor}
\usepackage{lineno,hyperref}

\usepackage{bm}
\usepackage{amsmath}
\usepackage{subfigure}

\usepackage{empheq}
\usepackage{gensymb}

\usepackage[utf8]{inputenc}
\usepackage[T1]{fontenc}

\usepackage{graphicx} 
\usepackage{graphicx,lineno,bm,float,soul,amsfonts,amsmath,siunitx}
\usepackage{xcolor,hyperref}
\usepackage{empheq}
\usepackage{multirow}
\ifdefined\pdfsuppresswarningpagegroup\pdfsuppresswarningpagegroup=1\fi

\newcommand{\B}[1]{{\bm #1}}

\usepackage{graphicx}
\usepackage{txfonts}

\ADsetup{
title    = {Designing 3D transfers: application to Earth-Moon},
author   = {A. K. de Almeida Jr$^{1,2}$\cor{}, T. Vaillant$^{2}$, J.P.F. Agostinho$^{3}$, A.F.B.A. Prado$^{4}$, L. B. T. Santos$^{5}$ \\
           },
address  = {
  1. Faculty of Sciences and Technology, University of the Azores, Rua da Mãe de Deus, 9500-321 Ponta Delgada, Portugal\sep
  2. Centro de Física da Universidade de Coimbra, Departamento de Física, Universidade de Coimbra, 3004-516 Coimbra, Portugal \sep
  3. Instituto de Geociências e Ciências Exatas, Universidade Estadual Paulista (UNESP), Rio Claro, São Paulo, Brazil \sep
  4. National Institute for Space Research, Av. dos Astronautas, 1758, São José dos Campos 12227–010, São Paulo, Brazil \sep
  5. Physics of Materials, Polytechnic School, University of Pernambuco, 50720-001 Recife, PE, Brazil
  },
email    = {allan.junior@uc.pt},
abstract = {
Three dimensional Earth-Moon transfers usually consume more fuel than planar ones, but are needed depending on the latitude of the launch site and the desired inclination of the final lunar orbit.
A systematic search of the optimal 3D transfers represents a true challenge due the dimension of the parameter set to span.
In this paper, we propose a new method to design 3D transfers using a tangential velocity constraint for the burns, modeled using the overdetermined constraints technique based on the Theory of Functional Connections for an optimal search of solutions in a numerically efficient way.
The developed method is applied here for a systematic study of optimal 3D Earth-Moon transfers, but could be used to design any type of 3D maneuvers.
  },
keywords = {
  Earth-to-Moon transfers \sep  Theory of Functional Connections \sep mission design \sep inclined orbits \\
},
}

\begin{document}

\maketitle  


\Nomenclature

\begin{center}
\begin{tabular}{|P{.45\linewidth}|P{.45\linewidth}|}
\hline
CRTBP & Circular Restricted Three-Body Problem \\
\hline
TFC & Theory of Functional Connections \\
\hline
\end{tabular}
\end{center}

\section{Introduction}

With Artemis mission, humanity enters in a new era of space exploration. A key problematic of space flights is the fuel quantity that must be loaded on board the spacecraft to perform the needed velocity changes (usually measured in km/s), imposing then constraints on the mass of payload and the power of the launcher. Since the Apollo era, progress have been made with new low-energy Earth-Moon transfers \cite{Belbruno1990,Belbruno2004} with a reduced fuel consumption in exchange for a longer transfer time (more than 30 days), and appropriated for lunar cargoes \cite{deAlmeida2026manifold}. Direct fast transfers (less than 10 days) like Apollo or Artemis missions are still needed for crewed missions. Such transfers usually consist in putting first the spacecraft into a parking orbit around Earth, and applying then an impulse to allow its transfer toward the Moon \cite{topputo2013optimal}.
As coplanar transfers need less fuel than out of plane transfers \cite{Lv2017,Zhang2021}, it is preferable to have the parking orbit in the Moon orbital plane to keep the transfer coplanar \cite{Penzo1963}.
However, depending on the launch site and on the desired final Lunar orbit, its is not always possible or optimal as it would necessitate an inclination change maneuver whose fuel cost proportional to the velocity of the spacecraft becomes prohibitively high from a 1$^{\circ}$ inclination change \cite{vallado7}.
Nonplanar direct Earth-Moon transfers are then usually investigated case-by-case depending on the launch site, the initial parking orbit, and the final lunar orbit \cite{Tselousova2019,Leonardi2025}.
Indeed, due to the complexity of designing three-dimensional transfers with existing methods, possible transfers are only investigated for parking orbits with a specific inclination, and no systematic studies have been performed yet similar to the planar case \cite{topputo2013optimal}.
As international space agencies and private initiatives increasingly pursue sustained lunar exploration, lunar gateways, and long-term cislunar infrastructure, physically consistent three-dimensional transfer frameworks may become increasingly relevant for future Earth–Moon transportation architectures.

Thus, in this paper, we perform a systematic study of 3D Earth-Moon transfers depending on the inclinations of the initial and final orbits.
For this, we develop a new method to design bi-impulsive 3D transfers, where both burns are tangential to the initial and final orbits and modeled as constraints in conjunction with the altitude of the burn point with respect to Earth (first burn) and Moon (second burn), which minimizes the associated costs \cite{vallado7,LAWDEN1962323,MARCHAL196991,MIELE200159,Pernicka1995,topputo2013optimal,QI2017106}.
The methodology is based on the Theory of Functional Connections (TFC) mathematical framework \cite{U-ToC}, allowing to search for the solution in a subspace where the constraints are satisfied. To this task, TFC derives a single expression - called the constrained functional - that exactly satisfies all the constraints of the problem.
Moreover, the recently developed overdetermined constraints technique allows to derive a constrained functional for constraints with nonlinear components based on a vector formulation of TFC \cite{tfc_segmentation}. This step is particularly important in our methodology, since the 3D transfer designed in paper is subject to multiple constraints with nonlinear components, as will be seen later in Secs. \ref{sec:parking} and \ref{sec:tangential}.
TFC has already been applied to design Earth-Moon planar transfers, solving the problem formulated as Two Point Boundary Value Problem \cite{fastTFC}, and was since then regularly perfected with the successive addition of the spherical coordinates \cite{deAlmeida2023}, a tangential velocity constraint \cite{akajtangential}, and subject to low thrust \cite{almeida24solar}. Besides the derivation of the overdetermined constraints, allowing to use TFC to nonlinear components constraints, the methodology shown in \cite{tfc_segmentation} segments the transfer in several parts, which allows us to vastly increase its numerical performance for designing complex transfers like the three-dimensional ones, and then to perform a systematic study of 3D Earth-Moon tranfers.

In this paper we evaluate the fuel equivalent costs $\Delta v$ associated to three-dimensional transfers, but also the connection between the inclination of the initial orbit around the Earth and its effect on the inclination of the arrival orbit around Moon. This is done for a bi-tangent impulsive maneuver strategy without inclination change maneuver.
We also include a 3D Patched Conics approach for comparison purpose.
The developed method was applied here to 3D Earth-Moon transfers, but could also be used to perform any type of bi-impulsive 3D transfers notably Earth-Mars transfers.
 
\section{Non-planar Earth-Moon transfer model}

\begin{figure}
\centering\includegraphics[scale=0.5]{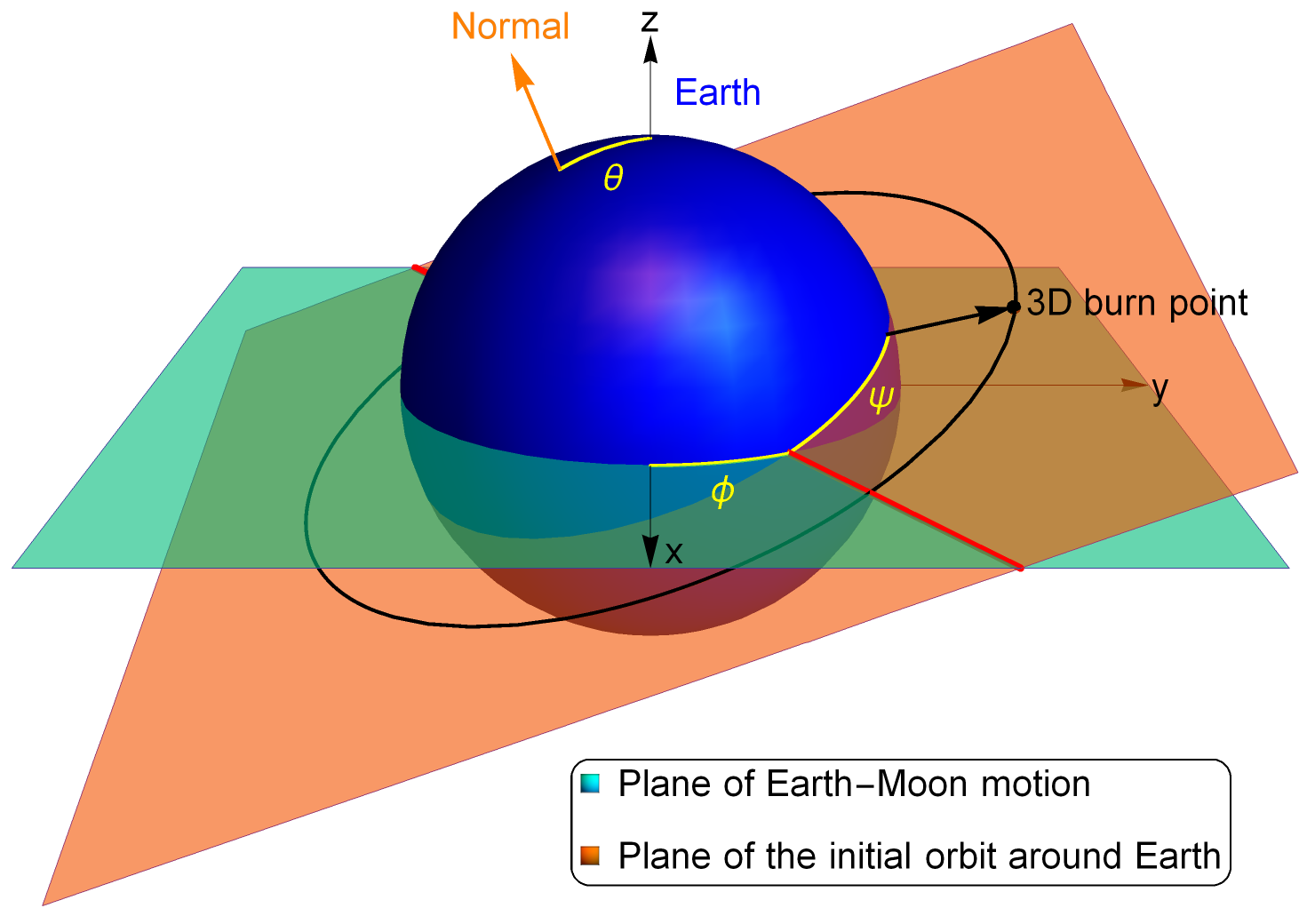}
\caption{The angles $\phi$, $\theta$ and $\psi$ in yellow connecting the plane of motion of Earth and Moon in green and the plane in orange of the initial inclined orbit around Earth in black. The line of nodes is in red. The trans-lunar injection of the 3D transfer is performed at the 3D burn point in a tangential direction with respect to this point in the orange plane.}
\label{fig:angles}
\end{figure}

We formulate the transfer considering the circular restricted three-body problem, where the Earth and Moon describe circular trajectories around their barycenter with a period of 27.32 days.
A spacecraft is in an initial circular parking orbit around Earth with an altitude of 167 km (such as those designed for the Apollo 11 mission in \cite{Berry1970}) and an orientation given by the Euler angles $\phi$, $\theta$ and $\psi$ (successively rotating around $z$, $x$, and $z$ axes), as shown in Fig. \ref{fig:angles} in black and yellow, respectively.
The angle $\theta$ is the inclination of the initial orbit with respect to the Moon orbital plane, and $\phi$ locates their intersection (line of nodes) with respect to the Earth-Moon axis.
The angle $\psi$ indicates the point of the initial orbit where a first impulse ($\Delta V_1$) is applied to transfer the spacecraft to the Moon.
A second impulse ($\Delta V_2$) is applied after a specified time of flight $T$ (ToF) when the spacecraft is close to the Moon in order to circularize the final orbit around it with an altitude of 100 km \cite{Berry1970}. A similar set of Euler angles is adopted to characterize the final circular orbit around Moon, with identical definition as those defined for the initial orbit around Earth shown in Fig. \ref{fig:angles}. They are denoted by $\phi_m$, $\theta_m$ and $\psi_m$, though.

As planar transfers usually consume less fuel than the three-dimensional ones, a common strategy is to perform the launch such that the initial parking orbit is on the Earth-Moon orbital plane - the case represented by $\theta=0$ in our formulation.
However, the Moon orbital plane, whose inclination with respect to the ecliptic is about $5^\circ$, precesses with a period of 18.6 years.
Thus, as the equator of the Earth has an inclination of about $23.5^\circ$ with respect to the ecliptic, the inclination of the Moon orbital plane with respect to the equator varies from $18.3^\circ$ to $28.6^\circ$ in a period of 18.6 years \cite{Roncoli2005,Williams2003}.
In case a spacecraft is launched from sites located at latitudes higher than this relative inclination (or lower than its negative value), the plane of the initial orbit around Earth cannot coincide with the Moon orbital plane, i.e. $\theta\ne0$, and thus a 3D maneuver is needed.

\subsection{Mathematical model}

In the CRTBP, Earth and Moon are fixed in the $x$ axis of a frame of reference centered in their barycenter and rotating around it with an angular velocity $\B{\omega}=\omega \hat{\B{z}}$, where $\hat{\B{z}}$ is a unit vector along the $z$ axis. The equation of motion of a spacecraft in this frame is \cite{symon}
\begin{equation}
	\label{eq:crtbp}
	\frac{\text{d}^2\B{r}}{\text{d}t^2} = -2\B{\omega} \times \frac{\text{d}\B{r}}{\text{d}t} - \B{\omega} \times \left(\B{\omega} \times \B{r} \right) - \frac{\mu_e}{r_{e}^{3}}  \B{r}_e - \frac{\mu_m}{r_{m}^{3}} \B{r}_m 
\end{equation}
where $\B{r}$ is the position vector of the spacecraft relative to the barycenter, $\B{r}_e$ and $\B{r}_m$ denote the position vectors of the spacecraft relative to the Earth and Moon, respectively, with norms $r_e$ and $r_m$. The gravitational parameters of the Earth and Moon are $\mu_e$ and $\mu_m$, respectively.
The distance of Earth and Moon from the center of the reference frame are $d_e$ and $d_m$, respectively, with $d_e=R\mu_m/(\mu_e+\mu_m)$ and $d_m=R\mu_e/(\mu_e+\mu_m)$, where $R$ is the distance between Earth and Moon.
The values of these parameters
are shown in Table \ref{tab:parameters} with the considered radii of Earth and Moon to compute the altitudes of the respective orbits.
\begin{table}[ht]
	\centering
	\begin{tabular}{lrl}
		\hline
		$R$ & $3.84405000 \times 10^{8}$ & m \\[0.5ex]
		$\mu_e$ & $3.975837768911438\times10^{14}$ & m$^3$/s$^2$  \\[0.5ex]
		$\mu_m$ & $4.890329364450684\times10^{12}$ & m$^3$/s$^2$ \\[0.5ex]
		$\omega$ & $2.66186135\times 10^{-6}$ & s$^{-1}$ \\[0.5ex]
		Radius of Earth&$6378 \times 10^{3}$ & m \\[0.5ex]
        Radius of Moon&$1738 \times 10^{3}$ & m \\[0.5ex]
		\hline
	\end{tabular}
    \caption{Values of the parameters for the Earth-Moon system \cite{topputo2013optimal,simo1995book,YAGASAKI2004313}.}
	\label{tab:parameters}
\end{table}


\subsection{\texorpdfstring{Physical representation of the angles $\phi$, $\theta$ and $\psi$}{Physical representation of the angles phi, theta and psi}}
\label{sec:angles}

The angle $\phi$ denotes the line of nodes representing the relative position of the Moon with respect to the plane of motion of the initial orbit. 
Due to the fact that the Moon is located along the $x$ axis of the rotating frame, this angle then varies with angular speed $\omega$ and is associated to the date to launch the mission - to start the transfer.

The angle $\theta$ represents the relative inclination of the plane of the initial orbit with respect to the plane of motion of Earth and Moon. Due to the obliquity of the Earth, the angle between the plane of its equator and the plane of the Earth-Moon motion varies from $18.3^\circ$ to $28.6^\circ$ \cite{Roncoli2005,Williams2003} in a repeating cycle of 18.6 years due to the period of regression of the longitude of the ascending node \cite{Roncoli2005}.
Missions departing from launching sites with latitudes higher than this relative inclination must consider an inclined arrival orbit as a function of the inclination of the initial orbit, as comprehensively investigated in this paper. It can be noted that an inclination change maneuver followed by a planar transfer would be prohibitively costly in this case due to the high velocities of the spacecraft orbiting the Earth in low altitudes.
Thus, we investigate here three-dimensional transfers that do not require the transfer to be done in the plane of the Earth-Moon motion. The costs associated to these three-dimensional transfers are then studied in this paper as functions of the relative inclination $\theta$. The connection between the inclinations of the initial orbit around Earth and the final orbit around Moon is also analyzed.


Lastly, the angle $\psi$ represents the specific point along the initial circular orbit around Earth such that the burn is applied in order to start the Trans Lunar Injection (TLI) impulsive maneuver.

As said before, a similar set of angles are defined to the final orbit around Moon, whose angles are denoted by $\phi_m$, $\theta_m$ and $\psi_m$. Thus, we have a total of 6 parameters defining the geometry of the transfer. It should be noted that the angles $\phi$ and $\psi$ associated to a given solution can be easily adapted to the mission: $\phi$ varies with the date to launch the mission and $\psi$ is a choice of the mission, denoting the position to apply the burn. On the other hand, $\theta$ cannot be modified without a costly inclination change maneuver and it thus influences the costs and the geometry of an Earth-Moon transfer. The angles $\phi_m$, $\theta_m$, and $\psi_m$ complete the transfer representing the geometric configuration of the final orbit around Moon.

\subsection{Initial and final circular parking orbits}
\label{sec:parking}


The Earth to Moon transfer problem is formulated as follows. The spacecraft is initially orbiting Earth in a circular parking orbit of altitude of $r_E=167$ km. 
This spacecraft is transferred to another circular orbit around Moon of altitude of $r_M=100$ km. 
The column vector $\B{r}_i$ denotes the position along the initial orbit around Earth where the first burn is applied. It is written as a function of the angles $\phi$, $\theta$ and $\psi$ using rectangular coordinates in the frame of reference rotating with the Earth and Moon as
\begin{equation}\label{eq:r0}
\B{r}_i = \begin{Bmatrix}
r_{E} \cos(\psi) \cos(\phi) - r_{E} \sin(\psi) \cos(\theta) \sin(\phi) - d_e \\
r_{E} \sin(\psi) \cos(\phi) \cos(\theta) + r_{E} \cos(\psi) \sin(\phi) \\
r_{E} \sin(\psi) \sin(\theta)
\end{Bmatrix}.
\end{equation}
Similarly, the final position of the spacecraft around Moon in the same frame of reference is given by
\begin{equation}\label{eq:rf}
\B{r}_f = \begin{Bmatrix}
r_{M} \cos(\psi_m) \cos(\phi_m) - r_{M} \sin(\psi_m) \cos(\theta_m) \sin(\phi_m) + d_m \\
r_{M} \sin(\psi_m) \cos(\phi_m) \cos(\theta_m) + r_{M} \cos(\psi_m) \sin(\phi_m) \\
r_{M} \sin(\psi_m) \sin(\theta_m)
\end{Bmatrix},
\end{equation}
where the indices $_m$ in the angles $\phi_m$, $\theta_m$ and $\psi_m$ are adopted to differentiate them from the departure angles $\phi$, $\theta$ and $\psi$ around Earth, although they are defined similarly with respect to the axes.

\subsection{Tangential initial and final velocities}
\label{sec:tangential}

A velocity impulse $\B{\Delta V}_1$ must be applied in a precise direction to guide an Earth orbiting spacecraft toward a target. 
For transfers between circular orbits, tangential impulses - applied either at the initiation or conclusion of the transfer - minimize overall maneuver costs \cite{vallado7}. This principle extends to transfers between elliptical orbits, provided the burns occur at the apsides \cite{LAWDEN1962323}. Within the analytical framework of the two-body problem, optimization procedures confirm that tangential velocities generally yield the most efficient trajectories \cite{MARCHAL196991}, though certain co-planar elliptical transfers may exploit geometric symmetries to achieve alternative optimal solutions.
In more complex dynamical models, such as those used for Earth-Moon or Moon-Earth transfers, numerical evidence demonstrates that tangential burns continue to approximately minimize impulsive maneuver costs \cite{MIELE200159}. Specifically, applying a tangential initial velocity at the departure point of an initial circular orbit maximizes fuel efficiency for Earth-to-Moon transfers \cite{Pernicka1995}. Because of its established efficiency and reliability, the tangential velocity assumption is frequently employed to solve complex transfer problems and evaluate trajectory costs, for instance, using the TFC \cite{akajtangential,tfcvariables} or multiple shooting \cite{topputo2013optimal, QI2017106} methods.

In this work, we constrain the initial velocity $\B{v}_i$ to be tangential to the initial circular parking orbit around Earth shown in black in Fig. \ref{fig:angles}. This is a very important step, because it decreases the dimensionality of the problem, implying in a much more efficient searching procedure in a subspace of the space of solutions where the costs are analytically optimized through the constrained functional, as will be shown in Sec. \ref{sec:overdetermined}. Using this constraint, the initial velocity can also be given as a function of the angles and a parameter representing its magnitude $v_i$ by
\begin{equation}\label{eq:v0}
\B{v}_i = v_i \begin{pmatrix}
- \sin(\psi) \cos(\phi) -  \cos(\psi) \cos(\theta) \sin(\phi) \\
 \cos(\psi) \cos(\phi) \cos(\theta) -  \sin(\psi) \sin(\phi) \\
 \cos(\psi) \sin(\theta)
\end{pmatrix}.
\end{equation}
Similarly, the final velocity around Moon is also constrained to be tangential to the orbit as function of the angles as
\begin{equation}\label{eq:vf}
\B{v}_f = v_f \begin{pmatrix}
- \sin(\psi_m) \cos(\phi_m) -  \cos(\psi_m) \cos(\theta_m) \sin(\phi_m) \\
 \cos(\psi_m) \cos(\phi_m) \cos(\theta_m) -  \sin(\psi_m) \sin(\phi_m) \\
 \cos(\psi_m) \sin(\theta_m)
\end{pmatrix}.
\end{equation}
The transfer is designed using a bi-impulsive maneuver in a time of flight $T$. The fuel equivalent $\Delta v$ cost is evaluated as the sum of the first impulse to leave the initial orbit around Earth plus the second impulse to enter the final orbit around Moon as $\Delta v=\|\B{\Delta V}_1\|+\| \B{\Delta V}_2\|$, where $\B{\Delta V}_1=\B{v}_i-\B{v}_e$ and $\B{\Delta V}_2=\B{v}_f-\B{v}_m$, where $\B{v}_e$ and $\B{v}_m$ are the velocities of the initial and final circular orbits around Earth and Moon, respectively.

\subsection{The overdetermined constraints technique based on TFC}
\label{sec:overdetermined}

The main purpose of this section is to derive a constrained functional that analytically satisfies the initial and final positions and velocities derived in previous sections. This is done using the Theory of Functional Connections (TFC) mathematical framework designed to perform linear functional interpolation \cite{U-ToC}. TFC can generate constrained functionals that are expressions analytically derived to satisfy a given set of linear constraints. These functionals transforms complex constrained problems into simpler, unconstrained ones, reducing the solution search space. Furthermore, we take advantage of the recently developed overdetermined constraints technique shown in \cite{tfc_segmentation} based on a vector formulation of TFC \cite{almeida24solar} to analytically embed multiple restrictions of the non-planar Earth-Moon spacecraft transfer problem proposed in this paper into the constrained functional. We then solve a new set of unconstrained differential equation derived based on this constrained functional.

We adopt the constrained functional based on the overdetermined constraints given by $\B{r}_i$, $\B{v}_i$, $\B{r}_f$ and $\B{v}_f$. This constrained functional can be derived as \cite{tfc_segmentation}
\begin{eqnarray}\label{eq:Ns1} \nonumber
        	\B{r}(t)&=&\B{g}(t)-\B{g}(0)+\B{r}_i+t \left(\B{v}_i-\dot{\B{g}}(0)\right)\\ 
            &&+t^2 \left(\frac{\dot{\B{g}}(T)+2 \dot{\B{g}}(0)-2 \B{v}_i-\B{v}_f}{T}-\frac{3 (\B{g}(T)-\B{g}(0)+\B{r}_i-\B{r}_f)}{T^2}\right)\\ \nonumber
            &&+t^3 \left(\frac{-\dot{\B{g}}(T)-\dot{\B{g}}(0)+\B{v}_i+\B{v}_f}{T^2}+\frac{2 (\B{g}(T)-\B{g}(0)+\B{r}_i-\B{r}_f)}{T^3}\right).
\end{eqnarray}
In this work, these overdetermined constraints are the initial and final positions and velocities given by Eqs. \eqref{eq:r0},  \eqref{eq:rf}, \eqref{eq:v0}, and \eqref{eq:vf}, respectively.
This step is very important, because this methodology given by the overdetermined constraints shown in \cite{tfc_segmentation} allows us to combine multiple unknowns into a single solution. On the other side, traditional techniques, such as multiple shooting methods, only solve the problem for a given set of parameters, that are varied independently.
In our case, it would mean modify the angles $(\phi,\theta,\psi,\phi_m,\theta_m,\psi_m)$,  and the velocities $v_i$, and $v_f$ one at a time on a 8-dimensional grid in a computationally costly and repeating procedure. Instead, TFC allows for an integrated numerical convergence of everyone of these parameters.

In this paper, we consider non-planar transfers depending on the angle $\theta$ (Sec. \ref{sec:angles}) that is the inclination of the plane of the initial orbit around Earth with respect to the plane of the Earth-Moon motion. For each simulation, we then impose a value for $\theta$ and maintain all the other angles and the initial and final velocities as free parameters. The convergence of the obtained solutions is then used to investigate the influence of $\theta$ on the other parameters notably the inclination of the final orbit $\theta_m$ and the fuel equivalent transfer cost that depends on $v_i$ and $v_f$.
Thus, we specify the angle $\theta$ and the time of flight $T$ and apply the numerical procedure described in Sec. \ref{sec:numerical} to find a solution of the Earth-Moon transfer problem by determining the angles $\phi$, $\psi$, $\phi_m$, $\theta_m$ and $\psi_m$ and the magnitude of the initial and final velocities $v_i$ and $v_f$.

\subsection{Numerical procedure}
\label{sec:numerical}

The constrained functional is an analytical expression containing the constraints and depending on the free function.
We present below a numerical optimization procedure to generate this free function such that it satisfies the equations of the dynamics. The generation of this free function completes the process, since the solution is then completely obtained through the constrained functional.

It is important to note that the constrained functional is obtained independently of the equations of the dynamics of the system. Thus, although we show applications to the second order differential equation given by Eq.~\eqref{eq:crtbp}, the technique proposed in this paper can also be applied (or easily extended) to other differential equations.

Wiht the constrained functional $\B{r}=\B{r}(\dot{\B{g}},\B{g},t)$ obtained in Eq.~(\ref{eq:Ns1}) for the specified constraints, the equations of motion (Eq.~\ref{eq:crtbp}) become
\begin{equation}\label{eq:unconst}
    \ddot{\B{g}}(t) - \B{a}'(\B{g}, \dot{\B{g}},  t) = \B{0},
\end{equation}
for the free function $\B{g}(t)$
\begin{equation}\label{eq:freefunction}
    \B{g}(t) = L \, \B{h} (\tau(t)),
\end{equation}
where $L$ is a constant matrix of unknown coefficients of dimension $3 \times (m - k + 1)$, $\B{h}(\tau)$ a $(m - k + 1) \times 1$ vector composed of orthogonal basis functions, such as Chebyshev polynomials of the first kind \cite{abramowitzstegun1972}, $m$ denotes the highest degree (i.e., truncation order) of the polynomial basis, and $k$ the number of constraints. The basis functions in $\B{h}$ are linearly independent of the support functions, resulting in a total of $m - k + 1$ terms.
As Chebyshev polynomials are defined on the interval $[-1:1]$, we perform the following change of variable to map the time interval $[0: T_s]$
\[
\tau = \frac{2t}{T_s} - 1,
\]
with $\tau\in[-1:1]$.
The time interval is discretized into $N + 1$ points using Chebyshev-Gauss-Lobatto nodes \cite{lanczos1988applied}
\begin{equation}\label{eq:distr}
    t_j - t_0 = \left(1 - \cos\left(\frac{j \pi}{N}\right)\right) \frac{T_s}{2}, \quad \text{for } j \in \llbracket 0 : N \rrbracket.
\end{equation}
By combining the free function representation from Eq.~(\ref{eq:freefunction}) with the time discretization in Eq.~(\ref{eq:distr}), the differential equation in Eq.~(\ref{eq:unconst}) can be transformed into the discrete system of equations
\begin{equation}\label{eq:eqmotion3}
    \B{a}''(L,\phi,\psi,\phi_m,\theta_m,\psi_m,v_i,v_f) = \B{0},
\end{equation}
where $\phi$,$\psi$,$\phi_m$,$\theta_m$,$\psi_m$,$v_i$,$v_f$ are the unknowns embedded in the constrained functional given by Eq.~\eqref{eq:Ns1} through $\B{r}_i$, $\B{r}_f$, $\B{v}_i$, and $\B{v}_f$ according to Eqs.~\eqref{eq:r0}, \eqref{eq:rf}, \eqref{eq:v0}, and \eqref{eq:vf}. Therefore, $\B{a}'': \mathbb{R}^{3 \times (m - k + 1) + 7} \to \mathbb{R}^{3 \times (N + 1)}$ represents the residual function evaluated at the collocation nodes.

The solution to the resulting system of $3 \times (N + 1)$ nonlinear equations with $3 \times (m - k + 1) + 7$ unknowns is obtained using a nonlinear least squares optimization method \cite{NDE} minimizing the norm of the residual matrix $\B{a}''$. The procedure to obtain the numerical results is coded in Python with the aid of the TFC module \cite{tfc2021github} combined with automatic differentiation \cite{10.1145/355586.364791}.

Once the optimal matrix $L$ and the parameters $\phi$,$\psi$,$\phi_m$,$\theta_m$,$\psi_m$,$v_i$,$v_f$ are found, the free function $\B{g}(t)$ can be obtained using Eq.~(\ref{eq:freefunction}). Substituting this free function back into the constrained functional in Eq.~(\ref{eq:Ns1}) yields the complete solution $\B{r}(t)$ of the original system of differential equations in Eq.~(\ref{eq:crtbp}), incorporating the specified form of the specific force $\B{a}(\B{r}, \dot{\B{r}}, t)$.

After convergence, we adopt a continuation method to find solutions for a different time of flight $T$.
We first find values of $\textbf{L}$ and $\phi$,$\psi$,$\phi_m$,$\theta_m$,$\psi_m$,$v_i$,$v_f$ given by $\textbf{L}'$ and $\phi'$,$\psi'$,$\phi_m'$,$\theta_m'$,$\psi_m'$,$v_i'$,$v_f'$, respectively, that are solutions to the transfer problem for the \textit{overdetermined constraints} in a time of flight $T$.
After, we use these values $\textbf{L}'$ and $\phi'$,$\psi'$,$\phi_m'$,$\theta_m'$,$\psi_m'$,$v_i'$,$v_f'$ as initial guesses for the unknown coefficients matrices $\textbf{L}$ and $\phi$,$\psi$,$\phi_m$,$\theta_m$,$\psi_m$,$v_i$,$v_f$ to solve the transfer problem in a time of flight $T+\delta T$ and inclination $\theta+\delta \theta$, where $\delta T$ and $\delta \theta$ are displacements in time and the inclination, respectively, small enough to allow convergence to the next optimal solution of $\textbf{L}$ and $\phi$,$\psi$,$\phi_m$,$\theta_m$,$\psi_m$,$v_i$,$v_f$. The values $\delta T=0.1$ days and $\delta \theta=0.00017\\mathrm{rad}$ were adopted to obtain the numerical results shown in this paper.
By repeating this procedure, we then obtain the coefficients $\phi$,$\psi$,$\phi_m$,$\theta_m$,$\psi_m$,$v_i$,$v_f$ for the desired transfer and time of flight, completing the solution. More details can also be seen also be seen in Sec. 3.3 of \cite{akaj_poincare}.

\section{Results}

We investigate transfers from an initial parking orbit around Earth of 167 km altitude to a final circular parking orbit around Moon with a distance of 100 km from its surface, as explained in Sec. \ref{sec:parking}.
Two types of transfers are possible: 1) counter-clockwise leading to a prograde final orbit around the Moon and 2) clockwise yielding to a retrograde one \cite{fastTFC}.
These types of solutions present in the planar case the lowest costs for short time transfers (less than 10 days). We then use the overdetermined constraints with TFC technique to embedded all the angles into the algorithm, except the angle $\theta$ representing the relative inclination of the initial orbit around Earth with respect to the plane of motion of the Earth and Moon. In this sense, among the six angles related to the orientations of the initial and final orbits and positions of burn ($\phi,\theta,\psi,\phi_m,\theta_m,\psi_m$), $\theta$ is the only angle to be imposed. The other ones are free to be varied according to the possible transfer solution obtained through our technique. 
To obtain nonplanar transfers from Earth to Moon, we then adopt a continuation procedure based on $\theta$, starting from $\theta=0^\circ$ for both the prograde and retrograde planar solutions. 
It can be noted that we call prograde solutions the families generated from the prograde planar one for $\theta=\theta_m=0^\circ$, although when $\theta$ increases it can lead to retrograde final orbits with $\theta_m\geq 90^\circ$, and inversely for the retrograde solutions. Thus, the words ``prograde'' and ``retrograde'' refer to the name of these families.

\subsection{\texorpdfstring{The $\Delta v$ cost as function of the inclination}{The Delta-v cost as function of the inclination}}

\begin{figure}
\centering
\includegraphics[scale=0.5]{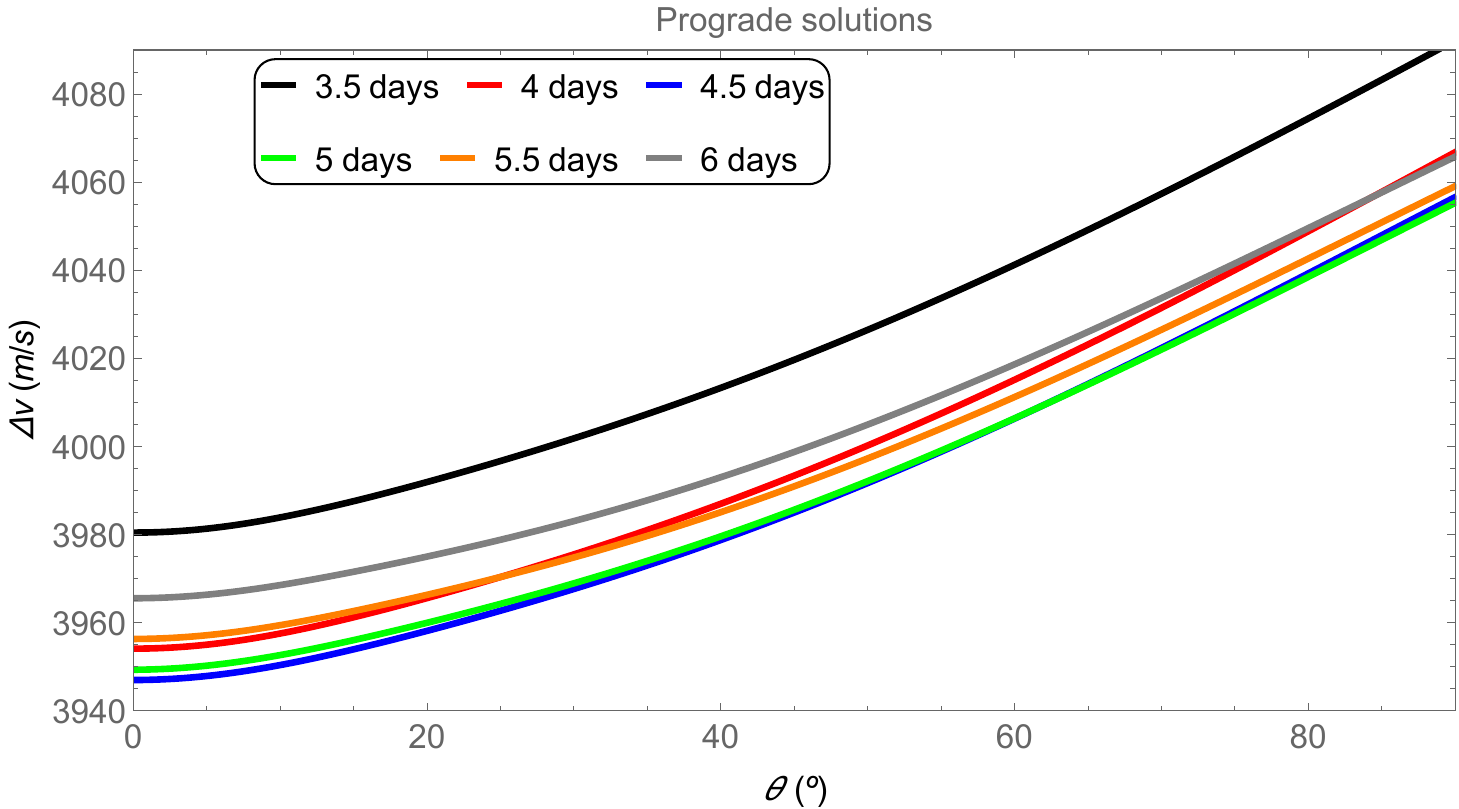}
\includegraphics[scale=0.5]{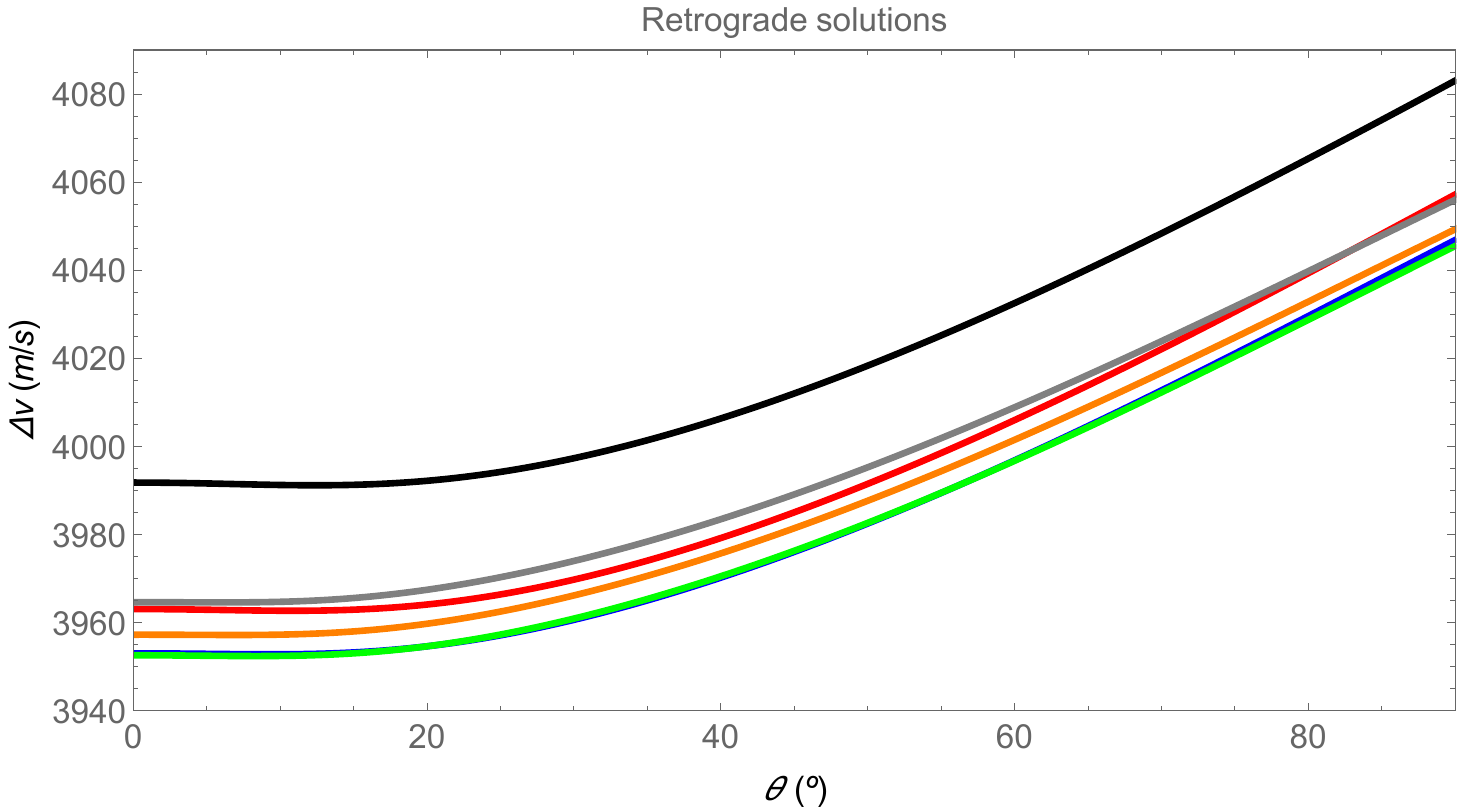}
\caption{The $\Delta v$ costs as functions of the inclination of the initial orbit $\theta$ for several values of times of flight between 3.5 and 6 days (in color) for prograde (above) and retrograde (below) solutions.}
\label{fig:dv_th_T1}
\end{figure}

The $\Delta v$ fuel equivalent cost to transfer a spacecraft from Earth to Moon is shown in Fig. \ref{fig:dv_th_T1} for several values of the inclinations of the initial parking orbit $\theta$ for the prograde (above) and retrograde (below) types of solutions. It can be noted that in the prograde case the lowest costs for small inclinations are represented by times of flight equal to 4.5 days, while for inclinations higher than 60$^\circ$, transfers with ToF=5 days become cheaper. The same tendency can be seen for solutions with ToF equal to 4 days (cheaper for lower inclinations) and 6 days (cheaper for inclinations close to polar orbits). Transfers with ToF equal to 4 days are cheaper than with ToF equal to 5.5 days for lower inclinations, but more expensive for higher inclinations.

\begin{figure}
\centering
\includegraphics[scale=0.5]{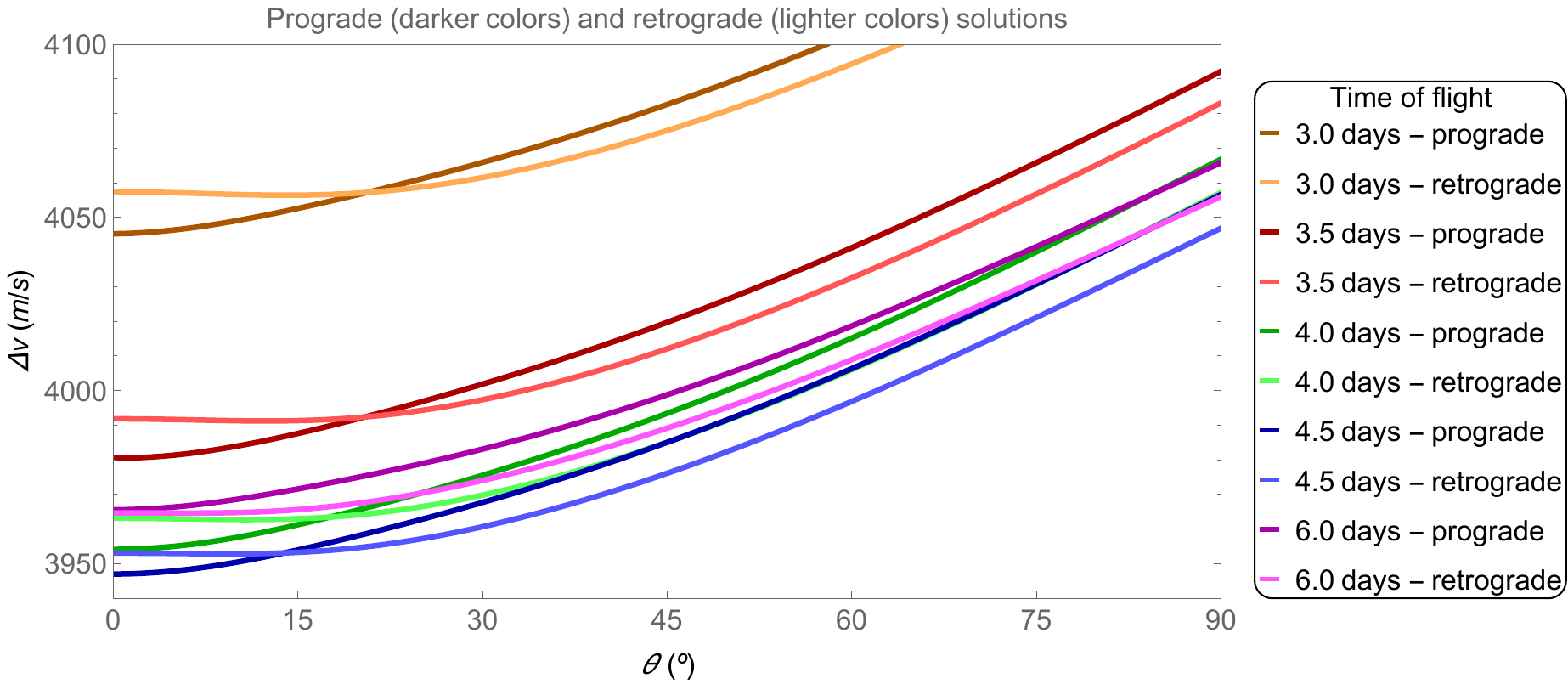}
\caption{Comparisons of the $\Delta v$ costs as functions of the inclination of the initial orbit $\theta$ for several values of times of flight between 3 and 6 days (in color) between prograde (in darker colors) and retrograde (in lighter colors) solutions.}
\label{fig:dv_th_T2}
\end{figure}

In fact, in general, the prograde solutions are cheaper than the retrograde solutions, but only for inclinations lower than a specific value, from where retrograde solutions become cheaper. This specific value of the inclination increases when we decrease the ToF. It can be 0$^\circ$ for ToF equal to 6 days, 15$^\circ$ for ToF close to 4.5 days and 20$^\circ$ for ToF close to 3 days. We can clearly see this pattern where both solutions cross each other in Fig. \ref{fig:dv_th_T2}. These difference can be up to about 10 m/s.

\begin{figure}
\centering
\includegraphics[scale=0.5]{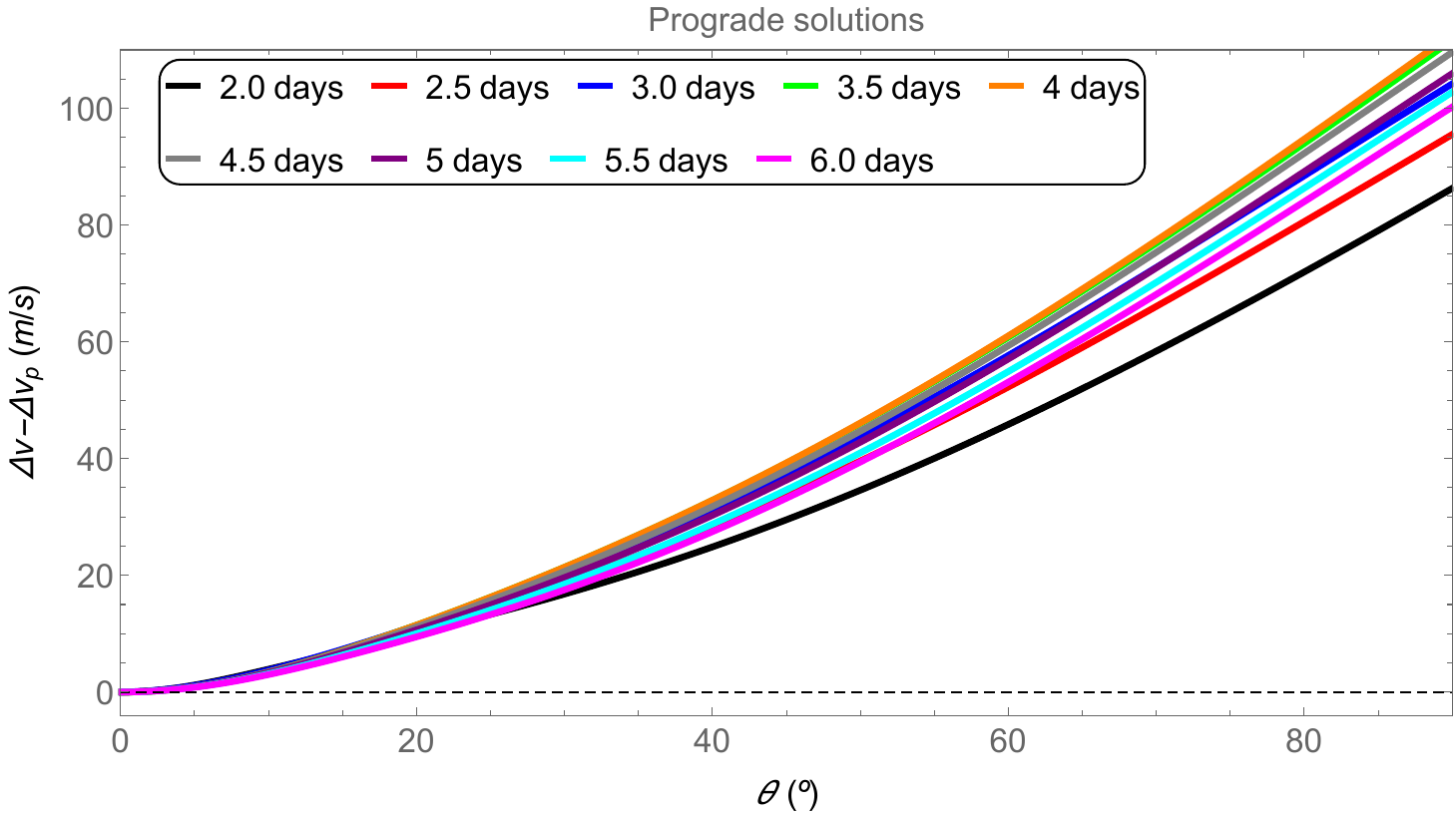}
\includegraphics[scale=0.5]{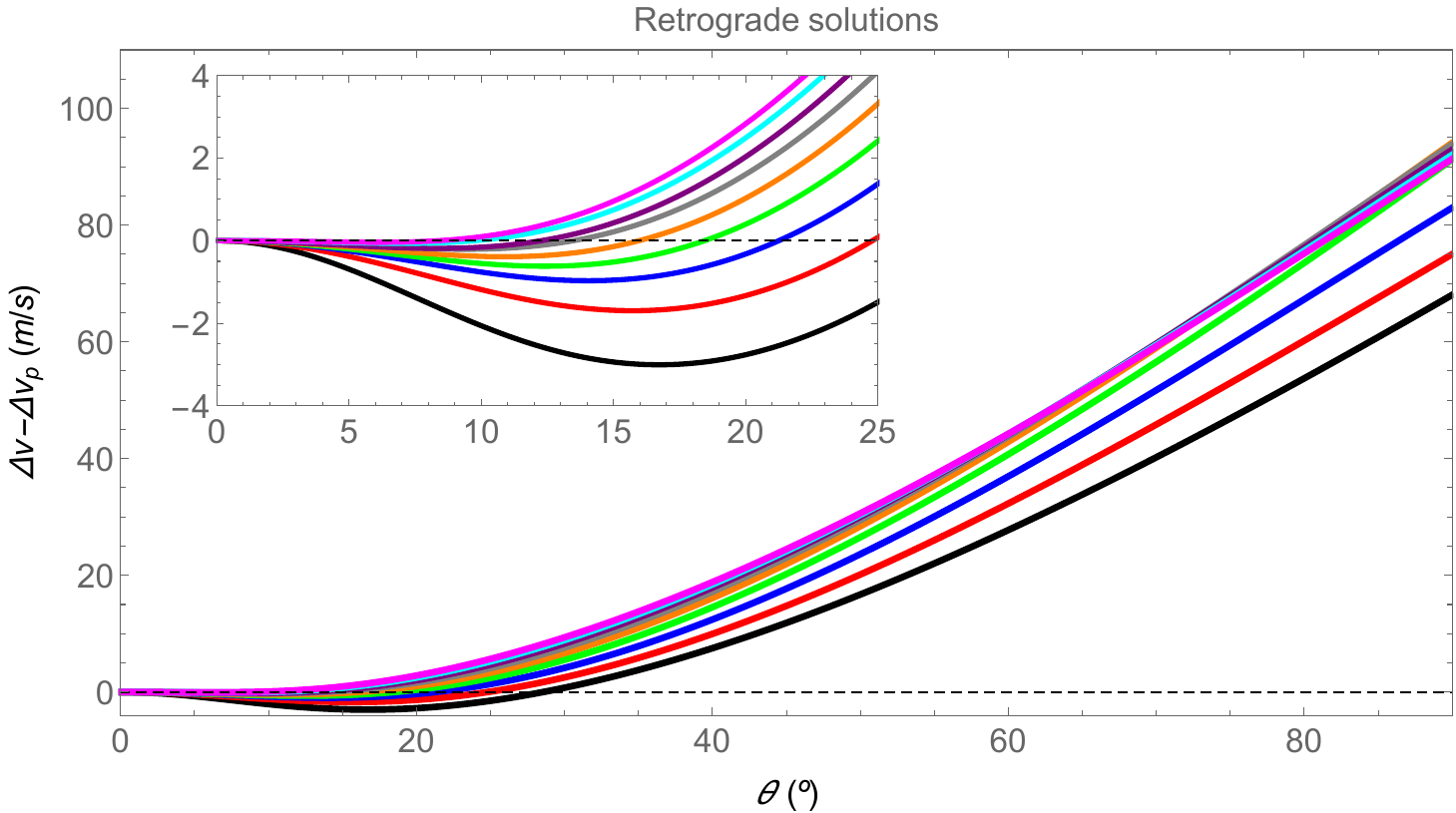}
\caption{The difference between the $\Delta v$ cost as function of the inclination of the initial orbit $\theta$ and the $\Delta v_p$ for planar transfers ($\theta = 0^\circ$) for several values of times of flight between 2 and 6 days (in color) for prograde (above) and retrograde (below) solutions.}
\label{fig:diff_dv1}
\end{figure}

In order to better visualize the costs, the difference $\Delta v-\Delta v_p$, where $\Delta v_p$ is the transfer cost $\Delta v$ evaluated for the planar case (evaluated for $\theta=0$), is shown in Fig. \ref{fig:diff_dv1} for both prograde (above) and retrograde (below) solutions, which indicates the increase of the cost with the inclination.
We can see in Fig. \ref{fig:diff_dv1} that the difference $\Delta v-\Delta v_p$ increase with the inclination.
We can also see in Fig. \ref{fig:diff_dv1} that - for the retrograde solutions - increasing the inclination can decrease the $\Delta v$ costs for lower inclinations (for instance up to 25$^\circ$ for ToF equal to 2.5 days), although the gains are lower than 4 m/s.
In this case, increasing the parking orbit inclination does not increase the transfer cost and can even decrease it for low inclinations.

\begin{figure}
\centering
\includegraphics[scale=0.5]{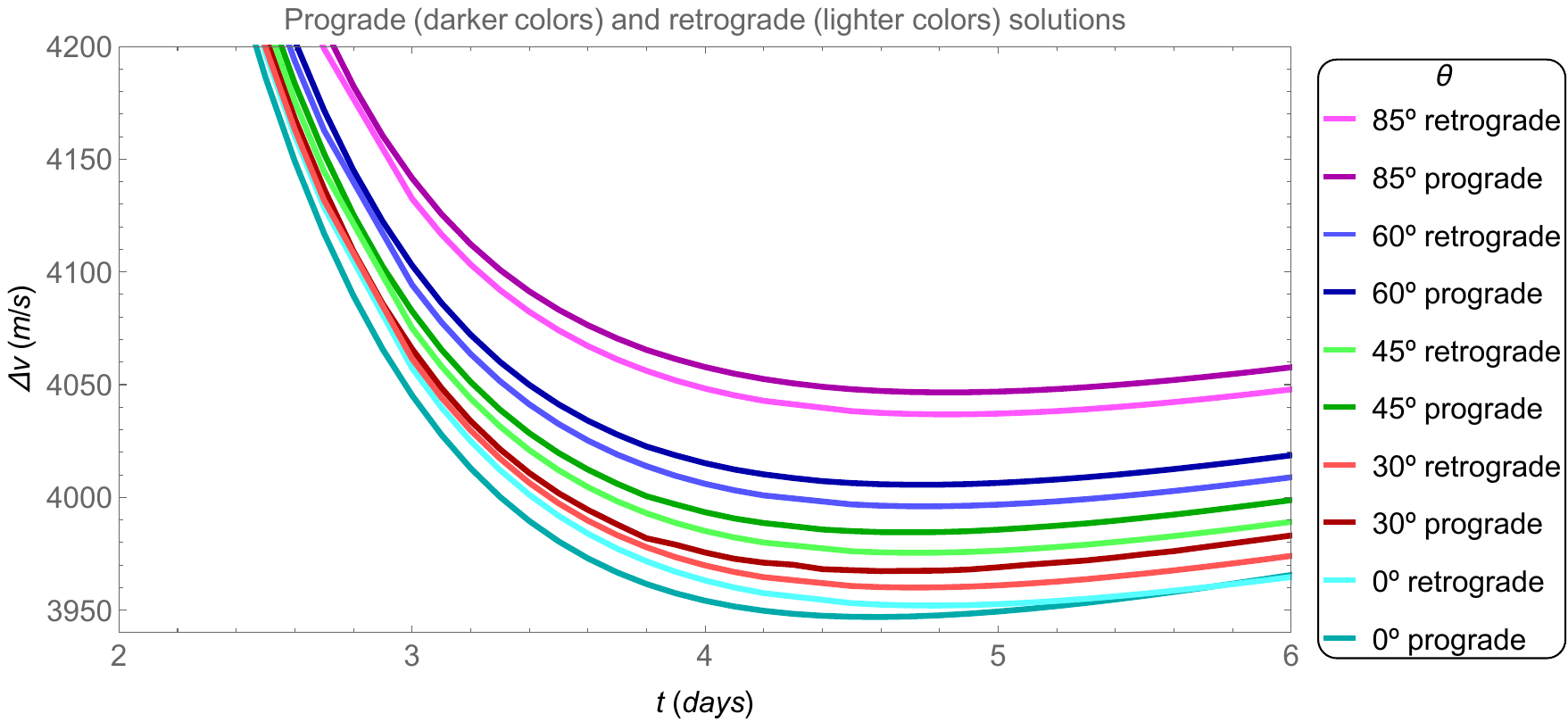}
\caption{The $\Delta v$ cost as function of the time of flight for several values of the inclination of the initial orbit $\theta$ and both prograde (darker colors) and retrograde (lighter colors) solutions.}
\label{fig:dvt}
\end{figure}

The total cost with respect to the ToF can be seen in Fig. \ref{fig:dvt} for several values of the inclination between 0$^\circ$ (the planar case) and 85$^\circ$.
Due to the obliquity of the Earth (23.5$^\circ$), a polar orbit around Earth has an inclination $\theta$ with respect to the Earth-Moon plane of motion between 61.5$^\circ$ up to 71.5$^\circ$, depending on the change of the inclination of the Moon with respect to the ecliptic (varying $\pm 5^\circ$).
Thus, in practical, the inclinations of the parking orbits around Earth are usually lower than these values. It can be noted from Fig. \ref{fig:dvt} that the Tof of the transfer associated to the minimum cost slightly increases with the inclination of the orbit, in the range of 4.5 days to 5 days.

\subsection{Crossing between inclinations of the initial and final orbits}

\begin{figure}
\centering
\includegraphics[scale=0.5]{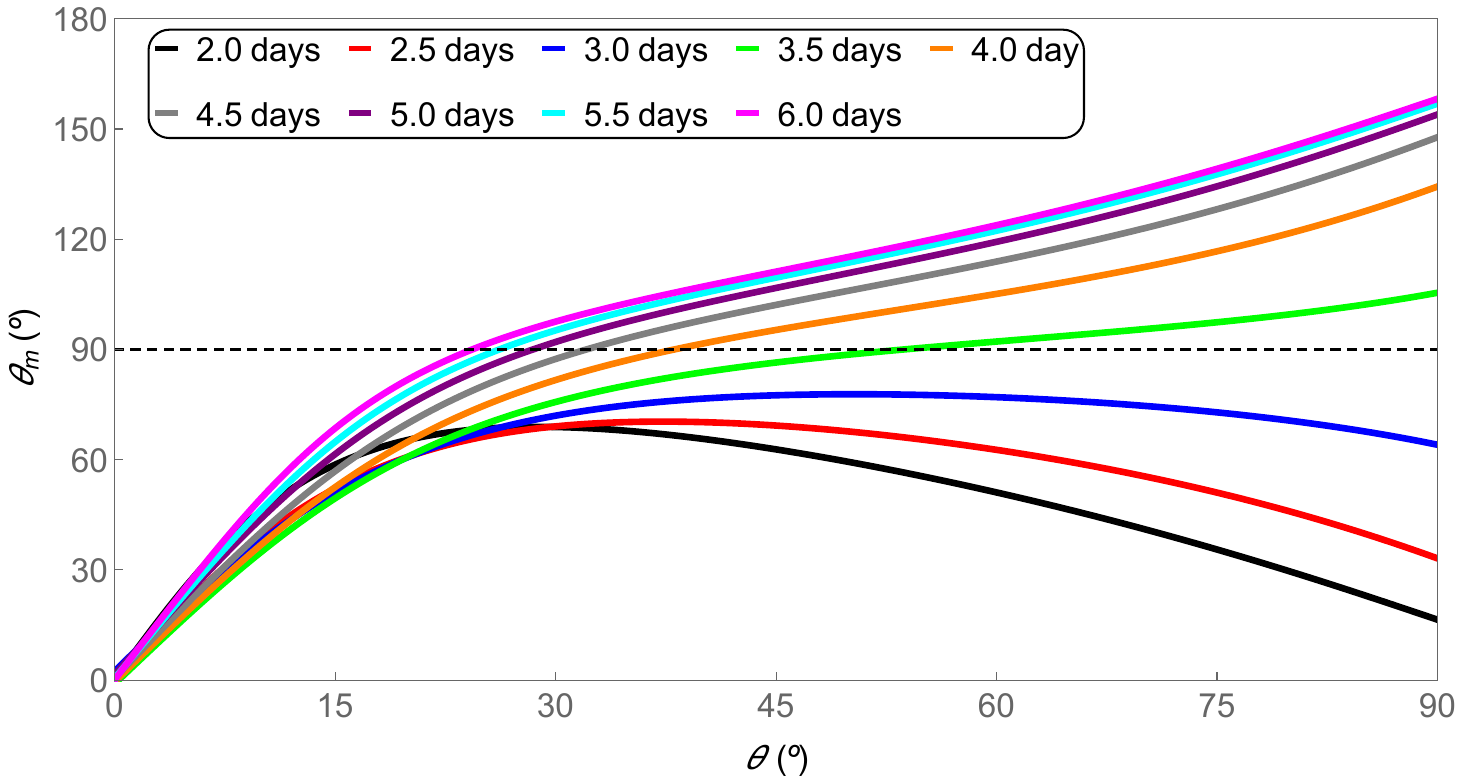}
\includegraphics[scale=0.5]{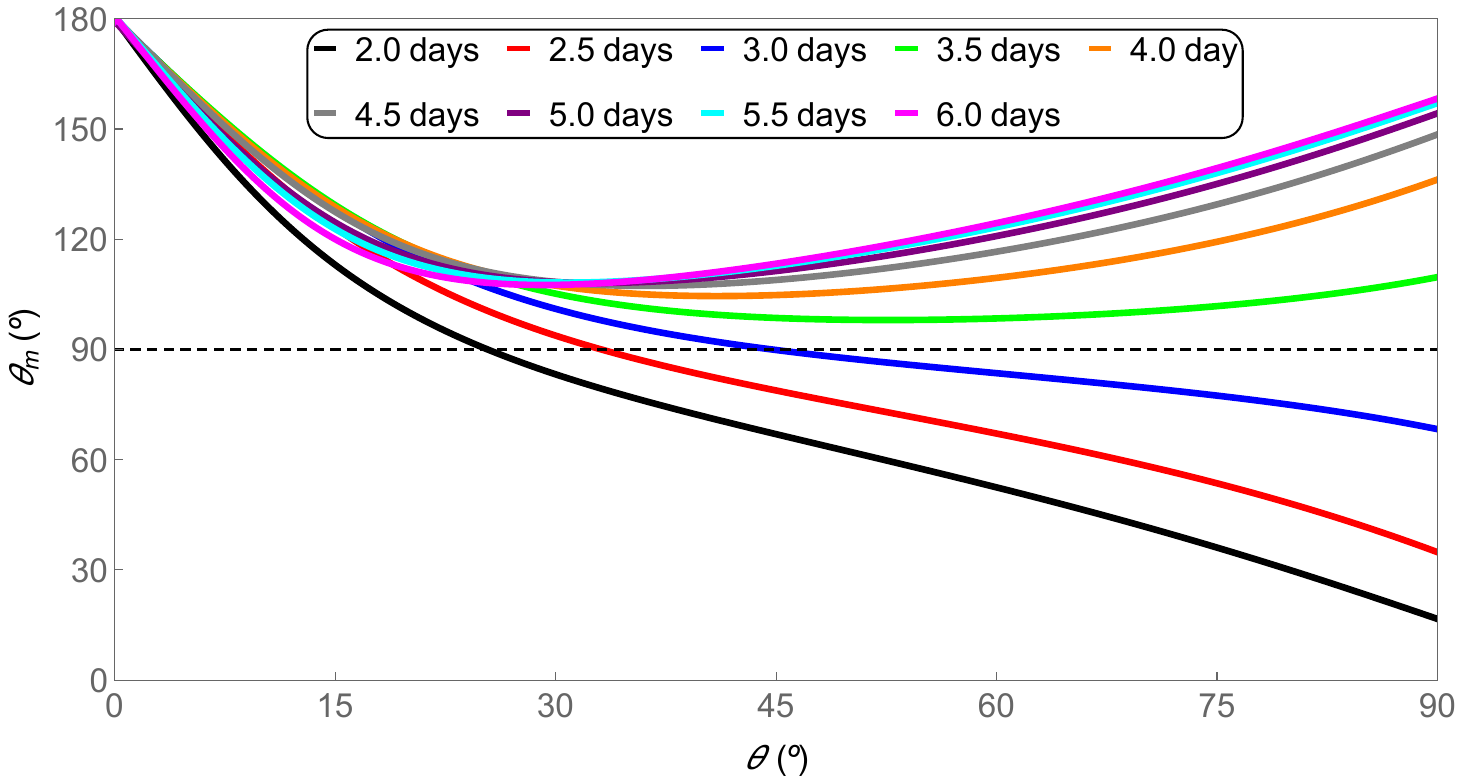}
\caption{The connections between the inclination of the initial orbit around Earth and inclination of the final orbit around Moon for prograde (above) and retrograde (below) solutions and several times of flight between 2 and 6 days.}
\label{fig:thm_th1}
\end{figure}

\begin{figure}
\centering
\includegraphics[scale=0.59]{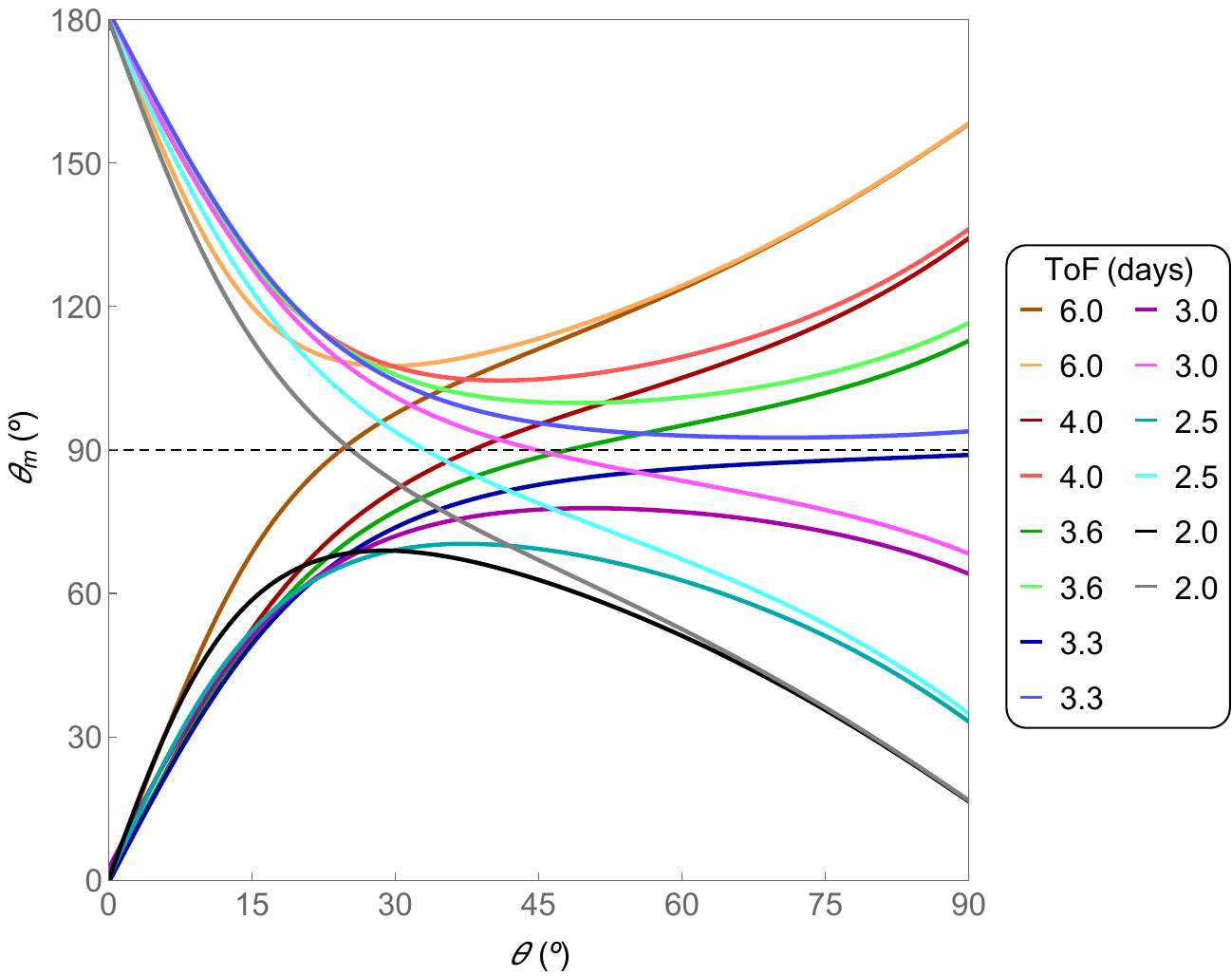}
\caption{The connections between the inclination of the initial orbit around Earth ($\theta$) and inclination of the final orbit around Moon ($\theta_m$) showing both prograde (darker colors) and retrograde (lighter colors) solutions for several times of flight between 2 and 6 days.}
\label{fig:thm_th2}
\end{figure}

In this section, we investigate the effects of changing the inclination of the initial orbit around Earth over the inclination of the final orbit around Moon. This relation is revealed in Fig. \ref{fig:thm_th1} for transfers with several values of ToF between 2 days and 6 days. The inclination of the final orbit around Moon is drastically influenced by the inclination of the initial orbit around Earth, but also by the ToF of the transfer. For instance, if a polar orbit of $\theta_m \approx 90^\circ \pm 6.5^\circ$ (considering the $5^\circ$ due to the inclination of the orbit and $1.5^\circ$ due to the obliquity of the Moon \cite{Colombo1966}) is desired by the mission, the prograde solutions are avaliable for ToF larger than 3.5 days, while retrograde solutions are available for ToF lower than 3 days, according to Fig. \ref{fig:thm_th1}.
Thus, we can observe in this figure that the value of the inclination of the final orbit depends on the time of flight and the inclination of the initial orbit.
Such information can be very useful for the early stage of a mission design. In another important example, in case the initial parking orbits around Earth is has 45$^\circ$ inclination ($\theta=45^\circ$) and the mission requires a final polar orbit around the Moon ($\theta_m=90^\circ$), the results contained in Fig. \ref{fig:thm_th1} show either two options: a ToF for the transfer equal to 3 days following the retrograde solution or a ToF between 3.5 and 4 days following the prograde solution.
In fact, we can see in Fig. \ref{fig:thm_th1} the inclination of the final orbit around Moon as function of the initial orbit around Earth (and the time of flight), for every one of the two types of solutions (prograde and retrograde). For lower values of the initial inclination $\theta$, varying the ToF does not significantly varies the final inclination $\theta_m$, but as the inclination $\theta$ increases, the effect of the ToF over the final inclination $\theta_m$ becomes more important. For instance, for an initial inclination $\theta=30^\circ$, the range of the final orbit is roughly within 60$^\circ$ to 100$^\circ$ for prograde solutions, depending on the ToF and within 80$^\circ$ to 110$^\circ$ for retrograde solutions, also depending on the ToF. Thus, for an initial orbit around Earth with inclination $\theta=30^\circ$, the range of the inclination of the final orbit around Moon is between 60$^\circ$ and 110$^\circ$, depending on the ToF and the type of solution (prograde or retrograde). No other value outside this range is allowed for these types of solutions. This range due to the crossing between prograde and retrograde types of solutions is clearly shown in Fig. \ref{fig:thm_th2}. Considering the departure from an initial orbit with inclination $\theta=15^\circ$, the two possibilities are to arrive in a final orbit with inclination in the ranges between 45$^\circ$ to 60$^\circ$ for prograde orbits and between 120$^\circ$ to 140$^\circ$ for retrograde orbits. When considering $\theta=45^\circ$, the range is slightly increased in comparison with the range for $\theta=30^\circ$ analyzed above. It can also be noted that both prograde and retrograde solutions tend to meet each other for higher inclinations $\theta$.



\subsection{Comparison with a Patched-Conics approach}

In order to have good initial guesses and numbers for comparison, we now compare to the well-known “Patched-Conics” Approach combined with the “Hohmann Transfer” in order to compute Earth-Moon transfers in an approximated form \cite{Chobotov1996}.
The idea is to divide the trajectories in two parts, both of them modeled by the “Two-Body” dynamics: an escape from Earth and an insertion into the Moon.

The first step is to get the semi-major axis of the Hohmann transfer, $a_t$,
\begin{equation}
a_t=\frac{r_p+r_a}{2},
\end{equation}
which is obtained from the perigee distance $r_p$ (the distance of the spacecraft to the center of Earth when it is in its initial orbit) and the apogee distance $r_a$ (the Earth-Moon distance).
Since the initial orbit around the Earth is inclined by the angle $\theta$, the whole transfer will be performed in this plane to avoid expensive plane change maneuver, and will be then in three dimensions.
Thus, new equations need to be derived to complement the vast literature that exist for planar maneuvers.

\begin{figure}
\centering
\includegraphics[scale=0.5]{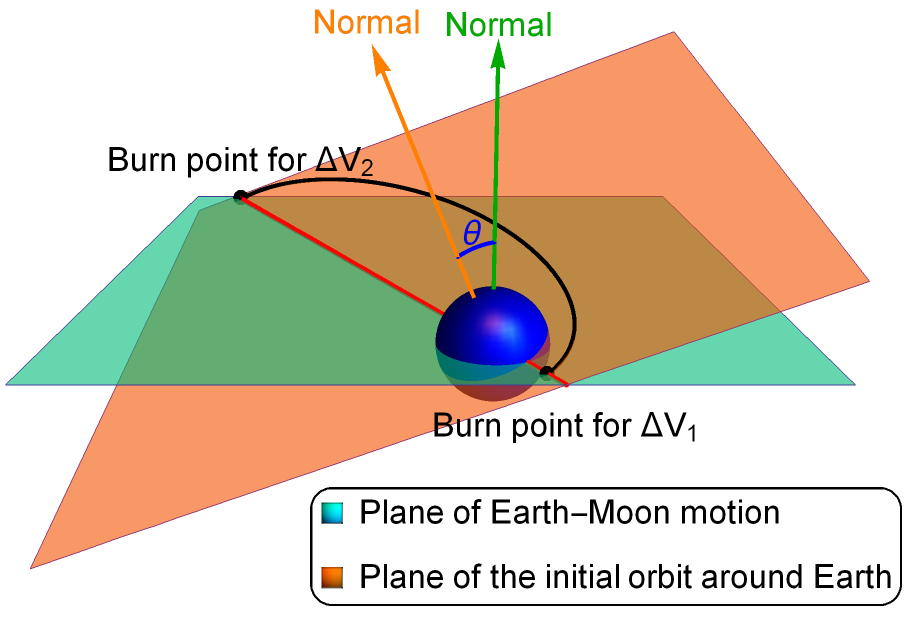}
\includegraphics[scale=0.56]{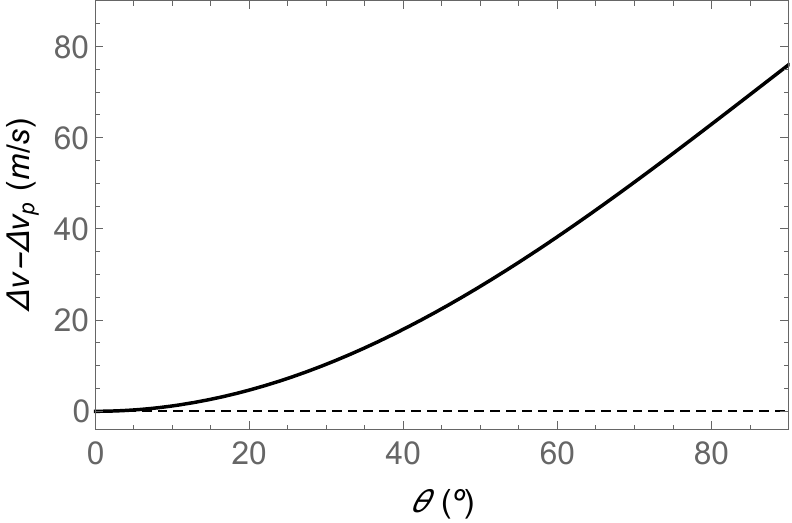}
\caption{Left side: geometry of the 3D “Patched Conics” Earth-Moon transfer. Right side: Patched conics based difference in the total impulse of the three dimensional transfer magnitude compared to the planar transfer (m/s) with respect to $\theta$.}
\label{fig:patchedconics_fig3}
\end{figure}

Compared with the more usual planar change, this maneuver will have the second impulse with a larger magnitude during the insertion of the spacecraft in the Moon, because the arrival velocity with respect to the Moon will have a component perpendicular to the orbital plane of the Moon, in addition to the planar component (left side of Fig. \ref{fig:patchedconics_fig3}).
This is a cost to be paid by leaving from an orbit around the Earth that is not coplanar with the orbital plane of the Moon.

The initial impulse for the maneuver is similar to the one for the planar Hohmann Transfer, and depends on the altitude of the initial orbit. The only difference from the planar transfer is that, since the impulse is applied in the direction of motion of the satellite and the initial orbit is not in the orbital plane of the Moon, this impulse and the initial velocity have a component perpendicular to the orbital plane of the Moon.
The second impulse aims to circularize the orbit around the Moon at any inclination.
For this, the arrival velocity $V_{am}$ with respect to the Moon considering the three dimensions is needed,
\begin{equation}
V_{am}=\sqrt{V_m^2+V_{at}^2-2V_mV_{at}\cos \theta},
\end{equation}
where $V_m$ is the velocity of the Moon with respect to the Earth and $V_{at}$ is the velocity of the spacecraft with respect to the Earth in the transfer orbit when it reaches the Moon.
After that, it is necessary to use the conservation of energy to find the velocity of the spacecraft with respect to the Moon at its perilune, $V_{pm}$, where the second impulse is applied to circularize the orbit
\begin{equation}
V_{pm}=\sqrt{V_{am}^2+2\frac{\mu_m}{r_f}},
\end{equation}
where $\mu_m$ is the gravitational parameter of the Moon and $r_f$ is the radius of the final orbit of the spacecraft around the Moon.
The magnitude of the second impulse is the difference between $V_{pm}$ and the circular velocity of the spacecraft around the Moon in its final orbit.

To compare with the previous results, we assume an initial circular orbit around the Earth of altitude 167 km and a final circular orbit around the Moon of altitude 100 km. The time of flight using the patched conic technique for this transfer is 5.02 days, and this value cannot be varied for the Patched Conics technique. Figure \ref{fig:patchedconics_fig3} shows the difference in the total impulse of the three dimensional transfer magnitude compared to the planar transfer with respect to $\theta$. The first impulse always has the magnitude of 3135.72 m/s.
We can observe that the increase of the difference with the inclination has the same order of magnitude than the one observed with our new method in Fig. \ref{fig:diff_dv1}.

\section*{Conclusions}

This work proposed  a new method to design three-dimensional transfers between two parking orbits around massive bodies. The overdetermined-constraints formulation under the Theory of Functional Connections mathematical framework was adopted to analytically embed the departure and arrival optimal conditions into the constrained functional. This includes tangential velocities, while allowing the most of the geometric parameters (angles) and velocity magnitudes to be determined simultaneously. Combined with a numerical continuation technique, this approach enabled a systematic investigation of families transfers from Earth to Moon originating from planar prograde and retrograde solutions as the departure inclination and time of flight were varied.

The results for the Earth to Moon transfer case show that departure inclination and transfer duration influence the total $(\Delta v)$, as expected and confirmed using a Patched-Conics approach. 
Among the investigated solutions, the time of flight associated with the lowest cost shifts from approximately 4.5 days toward 5 days as the departure inclination increases. The relative cost of the prograde and retrograde branches also changes with these parameters: prograde solutions are generally less costly at small inclinations, whereas retrograde solutions become preferable beyond a threshold that depends on the transfer duration. The differences between the branches reach approximately 10 m/s in the cases examined. Moreover, some retrograde transfers exhibit small cost reductions, around 2 to 3 m/s, relative to their planar counterparts, showing that a modest three-dimensional departure with inclinations bellow $25^\circ$g could be adopted with no need to necessarily increase the maneuver cost and could even decrease it.

A central finding is the dependence of the arrival-orbit inclination on both the departure-orbit inclination and the time of flight. With both inclinations measured relative to the Earth–Moon orbital plane, the computed families identify combinations that connect the initial and final circular orbits using only tangential departure and insertion impulses. For example, a departure inclination of $(45^\circ)$ can yield an arrival inclination of $(90^\circ)$ through either the retrograde branch with a flight time of approximately 3 days or the prograde branch with a flight time of approximately between 3.5 and 4 days. These relationships provide useful guidance for preliminary mission design by identifying compatible departure conditions, transfer durations, and arrival inclinations without a separate orbit inclination maneuver, which is usually very costly.

The three-dimensional patched-conics formulation provides a complementary approximation of how departure inclination affects the lunar arrival velocity and insertion cost. Together with the numerical transfer families, it offers a basis for preliminary cost estimates and the selection of candidate trajectories for subsequent refinement. The patched conics simpler model based results - although limited (e.g. the time of flight cannot be varied), and treats the departure geometry, transfer duration, and lunar arrival orbit as coupled mission-design choices.


The method developed in this paper is based on the constrained functional derived for the overdetermined constraints in Eq.~\eqref{eq:Ns1}, which does not depend on the model adopted to design the transfer. Due to this independency, the methodology can be then applied in a future work to assess the costs and persistence of the relationships in the inclinations of 3D transfers in an ephemeris-based model, incorporating relevant perturbations and finite-duration burns, and investigate additional transfer families for longer times of flight and operational constraints.
The method was applied here for the Earth-Moon system, but could be used to design any type of 3D maneuvers in others systems than the Earth-Moon one.

\section*{Acknowledgements}
We acknowledge support from the Foundation for Science and Technology (FCT) and by the European Regional Development Fund (ERDF) through the Innovation and Digital Transition Thematic Program (COMPETE 2030), of Portugal 2030, and by the European Union - Operation No. 14981 - COMPETE2030-FEDER-00860300 SPACE.
We acknowledge support from CFisUC UID/04564/2025 with DOI identifier 10.54499/\allowbreak UID/\allowbreak 04564/\allowbreak 2025. We also thank CNPq projects 443116/2023-7, 200489/2026-7 and 201893/2025-8, the Coordenação de Aperfeiçoamento de Pessoal de Nível Superior - Brasil (CAPES) - Finance Code 001 (process no. 88887.985142/2024-00), and São Paulo Research Foundation (FAPESP), under grant 2022/11783-5.

\section*{Contributions}

Conceptualization, supervision, methodology, software, and figures: AKAJ; Bibliographic review and generation of the first version of the manuscript: AKAJ and TV. Discussion of the results, review, and edition: AKAJ, TV, JPFA, AFBAP, and LBTS.


\bibliographystyle{elsarticle-num}
\bibliography{proposal_references}


\subsection*{Author biography}

\begin{biography}[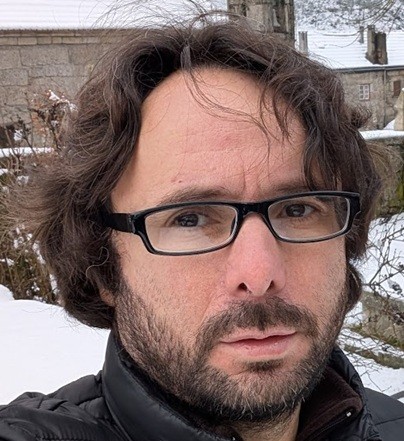]{A. K. de Almeida Jr.} I am a researcher in astrodynamics with experience in mission design since I obtained my PhD at INPE in Brazil in 2018. I am currently working at the University of Coimbra, Portugal. My expertise is developing novel methods and techniques to design orbit transfer / station keeping and perform satellite characterization / orbit determination. Email: \textit{allan.junior@uc.pt};
\end{biography}

\begin{biography}[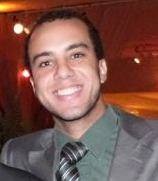]{L. B. T. Santos} Leonardo Barbosa Torres dos Santos received his B.Sc. degree in Physics from the Federal Rural University of Pernambuco (UFRPE), Brazil, in 2014, and his M.Sc. (2017) and Ph.D. (2021) degrees in Space Engineering and Technology from the National Institute for Space Research (INPE), Brazil. He was a visiting researcher at Politecnico di Milano, Italy, where he participated in the Lunar Meteoroid Impact Observer (LUMIO) mission project funded by the European Space Agency (ESA). His research interests include orbital dynamics, nonlinear dynamics, celestial mechanics, and trajectory design for spacecraft and small body missions. He is currently a professor at the Polytechnic School of Pernambuco, University of Pernambuco (UPE), and serves as a reviewer for Advances in Space Research and Planetary and Space Science. Dr. Santos received an Honorable Mention at the INPE Institutional Training Program in 2022. Email: \textit{leonardo.santos@upe.br};
\end{biography}

Emails of other authors: \textit{vaillant@ua.pt} (T. Vaillant);
\textit{j.agostinho@unesp.br} (J. Agostinho);
\textit{antonio.prado@inpe.br} (A. Prado);

\end{document}